\documentclass[aps,prl,twocolumn,superscriptaddress]{revtex4-2}

\usepackage{amsthm}
\usepackage{amsmath,bm}
\usepackage{amssymb}
\usepackage{amsfonts}
\usepackage{graphicx}
\usepackage{txfonts}
\usepackage{xcolor}
\usepackage{float}
\usepackage{braket}
\usepackage[colorlinks=true,linkcolor=blue,citecolor=blue,urlcolor=blue]{hyperref}
\usepackage[capitalise]{cleveref}
\usepackage{bbm}
\usepackage{changes}

\newcommand{\ignore}[1]{} 

\newcounter{SaveEqnCntr}

\begin{document}

\title{Non-Gaussian Entanglement Hierarchy Based on the Schmidt Number}

\author{Jiajie Guo}
\affiliation{State Key Laboratory of Artificial Microstructure and Mesoscopic Physics, School of Physics, Frontiers Science Center for Nano-optoelectronics, $\&$ Collaborative Innovation Center of Quantum Matter, Peking University, Beijing 100871, China}

\author{Shuheng Liu}
\affiliation{State Key Laboratory of Artificial Microstructure and Mesoscopic Physics, School of Physics, Frontiers Science Center for Nano-optoelectronics, $\&$ Collaborative Innovation Center of Quantum Matter, Peking University, Beijing 100871, China}

\author{Matteo Fadel}
\email{fadelm@phys.ethz.ch}
\affiliation{Department of Physics, ETH Z\"{urich}, 8093 Z\"{urich}, Switzerland}

\author{Qiongyi He}
\email{qiongyihe@pku.edu.cn}
\affiliation{State Key Laboratory of Artificial Microstructure and Mesoscopic Physics, School of Physics, Frontiers Science Center for Nano-optoelectronics, $\&$ Collaborative Innovation Center of Quantum Matter, Peking University, Beijing 100871, China}
\affiliation{Collaborative Innovation Center of Extreme Optics, Shanxi University, Taiyuan, Shanxi 030006, China}
\affiliation{Hefei National Laboratory, Hefei 230088, China}

\begin{abstract}
Non-Gaussian entanglement is a promising resource in various quantum tasks.
A recently defined class identifies entanglement that cannot be generated by applying Gaussian operations to separable inputs. 
To further explore the entanglement in this context, we introduce a quantitative witness $E_{\rm NG}$ in bipartite bosonic systems, which satisfies $E_{\rm NG}=1$ for all Gaussian-entanglable states, while $E_{\rm NG}>1$ certifies non-Gaussian entanglement. 
Its ceiling $d=\lceil E_{\rm NG}\rceil$ provides a lower bound on the Schmidt number irreducible by Gaussian transformations, thereby defining a natural hierarchy of non-Gaussian entanglement. 
For pure states, the condition is sharp and the hierarchy reflects the complexity of state learning.
We benchmark the framework with some paradigmatic non-Gaussian states, such as NOON states and squeezed Kerr states, and analyze its robustness against loss. Moreover, we construct an experimentally economical NOON-type witness requiring only four density-matrix element measurements. 
These results establish an operationally meaningful and experimentally accessible framework for identifying non-Gaussian entanglement resources in continuous-variable quantum platforms.
\end{abstract}

\maketitle

\begin{figure}[t]
    \centering
    \includegraphics[width=.48\textwidth]{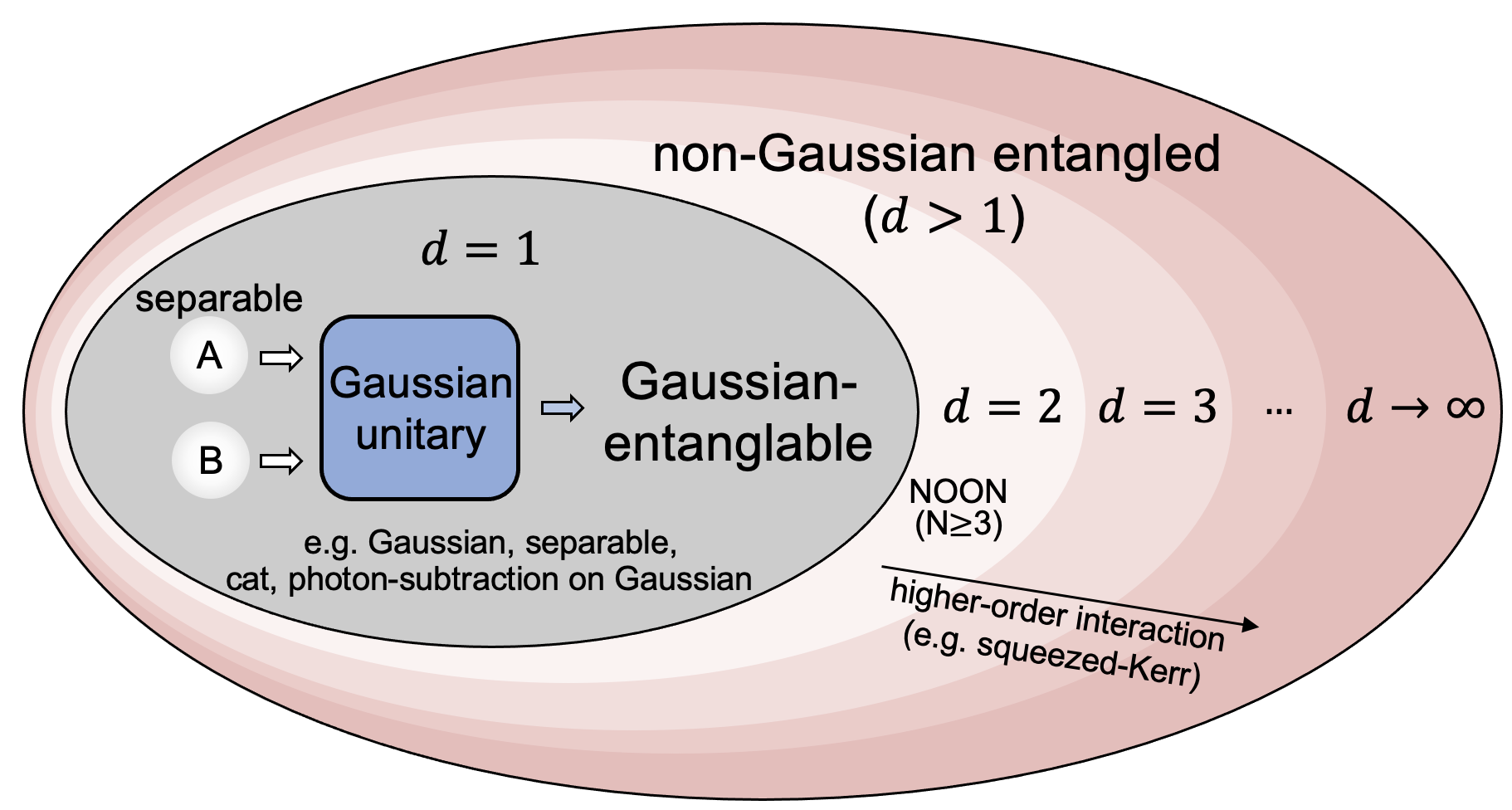}
    \caption{ \textbf{Hierarchy of Gaussian-entanglable and non-Gaussian entangled states.} The witness $E_{\text{NG}}(\rho)$ defines a hierarchy by a positive integer $d= \lceil E_{\text{NG}}(\rho) \rceil$. It satisfies $d=1$ for Gaussian-entanglable states $\rho \in \mathcal{GE}$, generated by applying a two-mode Gaussian unitary to separable inputs.   
    This class includes all Gaussian and separable states, as well as some non-Gaussian states such as two-mode cat states and photon subtractions on pure Gaussian states. 
    States with $E_{\rm NG}(\rho)>1$ are certified as non-Gaussian entangled, and larger $d$ witnesses higher non-Gaussian entanglement dimensions. For pure states, $E_{\rm NG}(\ket{\psi})>1$ is a sufficient and necessary condition for certifying genuine non-Gaussian entanglement, and $d=\lceil E_{\text{NG}}(\ket{\psi}) \rceil$ provides a lower bound on state learning complexity.   
    }
    \label{Fig1_IllustractionNGE}
\end{figure}

Entanglement is a key resource for quantum  communication, computation, sensing, and error correction~\cite{Horodecki2009, Gisin2007, Gottesman2001}.
In bosonic systems, such as optical modes, microwave cavities, and mechanical oscillators, entanglement takes on a particularly rich structure, owing to the natural division of states and operations into Gaussian and non-Gaussian classes~\cite{Weedbrook2012, Serafini2017, Walschaers2021}.
Although Gaussian operations can generate entanglement from separable inputs, they are fundamentally limited in the type and depth of entanglement they can create.
Indeed, Gaussian resources alone cannot achieve entanglement distillation~\cite{EisertPRL2002, Giedke2002, Fiurasek2002}, loophole-free Bell inequality violations~\cite{Banaszek1998}, or universal quantum computation~\cite{Lloyd1999, Bartlett2002}, making non-Gaussian entanglement an indispensable ingredient for next-generation bosonic quantum technologies.

Characterizing entanglement beyond the Gaussian regime is substantially more intricate.
For Gaussian states, the covariance matrix encodes all entanglement properties, and well-established criteria, such as the Peres-Horodecki positive partial transpose (PPT) condition~\cite{Peres1996, Horodecki1997}, Simon ~\cite{Simon2000} and Duan \textit{et al.} criterion~\cite{DuanPRL2000}, and the logarithmic negativity~\cite{Vidal2002, Adesso2004}, 
provide necessary and sufficient conditions for detection and quantification.
For non-Gaussian states, however, second-moment-based tools are generally incomplete: entanglement may reside in higher-order correlations and thus require information beyond the covariance matrix, such as witnesses involving higher-order moments~\cite{Shchukin2005,HilleryPRL2006}, phase-space criteria~\cite{LinPRL2023,LinPRL2024,LiuQuantum2026,LiuArXiv2026} or full state tomography~\cite{Peres1996,Horodecki1997,Vidal2002}.
In the finite-dimensional setting, the Schmidt number~\cite{Terhal2000, Sanpera2001} provides a hierarchy of entanglement dimension with direct operational meaning for quantum information tasks~\cite{Huber2013, Friis2019}.
Extending an analogous quantitative hierarchy to non-Gaussian bosonic entanglement has remained an open challenge.

A fundamental distinction in the entanglement of bosonic systems arises between states whose entanglement is Gaussian-generatable, i.e., producible by Gaussian transformations acting on separable inputs, and those for which no such protocol suffices, regardless of the non-Gaussianity of the input states~\cite{Zhao2025NatComm}.
The former class, the Gaussian-entanglable (GE) states, is operationally rich: it encompasses all Gaussian states, the outputs of boson sampling circuits~\cite{Aaronson2011, Madsen2022}, and multi-mode Gottesman-Kitaev-Preskill states~\cite{Gottesman2001, Larsen2025}. However, it remains strictly limited: For instance, NOON states $(|N,0\rangle + |0,N\rangle)/\sqrt{2}$ with $N \geq 3$, widely recognized as optimal metrological probes~\cite{Boto2000, Mitchell2004, Nagata2007}, cannot be deterministically generated by any Gaussian protocol and thus represent genuine non-Gaussian entangled (GNGE) states~\cite{Zhao2025NatComm, MattiaArXiv2026}.
While this distinction is now established, the derivation of criteria and experimentally practical witnesses capable of detecting non-Gaussian entanglement remains largely unexplored.
In particular, a quantitative characterization of the non-Gaussian entanglement structure, one that assigns a meaningful dimension to the entanglement irreducible by Gaussian unitaries and connects directly to observable quantities, is still lacking.

In this Letter, we introduce a witness $E_{\rm NG}$ for non-Gaussian entanglement characterization.
This quantity generalizes the standard negativity-based entanglement witness~\cite{Vidal2002} to the non-Gaussian setting. 
It satisfies $E_{\rm NG} = 1$ for all GE states, while $E_{\rm NG} > 1$ certifies non-Gaussian entanglement.
Its ceiling $d \coloneqq \lceil E_{\rm NG} \rceil$ witnesses the minimum Schmidt number that remains after Gaussian transformations, thereby inducing a hierarchy of non-Gaussian entanglement dimensions.
For pure states, the criterion is sharp and the hierarchy is connected to state-learning complexity. 
We benchmark our framework on some paradigmatic non-Gaussian states, and further evaluate its robustness against loss. 
Finally, to circumvent the cost of full state tomography, we derive a practical witness tailored to NOON-type states requiring only four density-matrix element measurements.

\vspace{2mm}
\textbf{Gaussian-entanglable and non-Gaussian entangled states.--} 
In a bipartite scenario, a pure state of two bosonic modes is called Gaussian-entanglable (GE) if it can be generated from a product state by a two-mode Gaussian unitary~\cite{Zhao2025NatComm}, 
\begin{align}\label{eq:pureGE}
|\psi_{\text{GE}} \rangle &= \hat{U}_{\text{G}}\ |\psi_A \rangle \otimes |\psi_B \rangle.
\end{align}
Crucially, the input states $|\psi_{A} \rangle$ and $|\psi_{B} \rangle$ are not necessarily Gaussian, for instance, they could be Fock states.
As an example, the state $(\ket{02}+\ket{20})/\sqrt{2}$ is GE~\cite{Zhao2025NatComm}, since it can be prepared from the product state $\ket{1}\otimes\ket{1}$ through Hong-Ou-Mandel interference~\cite{Hong1987} at a balanced beam splitter.

According to the Bloch-Messiah decomposition~\cite{Weedbrook2012}, any two-mode Gaussian unitary can be expressed as
\begin{align}
\hat{U}_\text{G} &= \left(\hat{D}_A \otimes \hat{D}_B \right) \hat{V}_{\text{II}} \left( \hat{S}_A \otimes \hat{S}_B \right) \hat{V}_{\text{I}},
\end{align}
where $\hat{D}_{\alpha}$ and $\hat{S}_\alpha$ are local displacement and squeezing operators acting on mode $\alpha\in\{A,B\}$, while $\hat{V}_{\text{I}} $ and $ \hat{V}_{\text{II}}$ are passive linear-optical interferometers. 
We denote the set of all two-mode Gaussian unitaries by $\mathcal{G}_2$. A pure state that cannot be expressed in the form of Eq.~\eqref{eq:pureGE} is called a \textit{pure genuine non-Gaussian entangled state}.

More generally, for mixed states, we consider the action of a two-mode Gaussian unitary $\hat{U}_{\text{G}}$ on a separable input state $\rho_{\text{sep}} = \sum_{i} c_i \rho_A^{(i)} \otimes \rho_B^{(i)} $, with $c_i\geq 0,\sum_i c_i =1$. 
This defines a class of GE states
\begin{equation}
    \mathcal{GE}=\left\{ \hat{U}_{\text{G}} \rho_{\text{sep}} \hat{U}_{\text{G}}^\dagger : \hat{U}_{\text{G}}\in \mathcal{G}_2,\ \rho_{\text{sep}}\;\text{is separable} \right\} ,
\end{equation}
such that a state outside this set, $\rho_{\text{NGE}} \notin \mathcal{GE}$, is called \textit{non-Gaussian entangled} (NGE).

A stronger notion is obtained by allowing classical mixture of GE states generated by different Gaussian unitaries. This defines the convex set
\begin{align}
\mathcal{GE}_{\rm mix} = \left\{ \sum_j p_j\ \rho_{\text{GE}}^{(j)} : p_j\ge0,\  \sum_jp_j=1,\
\rho_{\text{GE}}^{(j)}\in \mathcal{GE} \right\}.
\label{eq:rhoGE}
\end{align}
A state outside this set, $\rho_{\text{GNGE}} \notin \mathcal{GE}_{\rm mix}$, is called \textit{genuine non-Gaussian entangled} (GNGE). Clearly, $\mathcal{GE} \subseteq \mathcal{GE}_{\text{mix}}$, showing that GNGE is a stricter condition than NGE. In the following, we are interested in investigating approaches to detect both NGE and GNGE states.

\vspace{2mm}
\textbf{Certifying non-Gaussian entanglement.--}
The certification and characterization of entanglement in continuous-variable (CV) systems is central to several quantum-information tasks~\cite{BraunsteinRMP2005}. One powerful strategy is based on the Schmidt number~\cite{Terhal2000, Sanpera2001}, which provides a paradigmatic hierarchy of entanglement.
For a pure two-mode state with Schmidt decomposition $|\psi\rangle = \sum_i \lambda_i |n_i \rangle |m_i\rangle $, the Schmidt rank $\mathcal{SN}(|\psi\rangle)$ is the number of nonzero Schmidt coefficients. For a mixed state $\rho$, the Schmidt number is defined as the minimum Schmidt rank over all possible pure-state decompositions, 
\begin{align}\label{eq:DefSNmix}
\mathcal{SN} := \inf_{\mathcal{D}(\rho) } \max_{|\psi_i \rangle \in \mathcal{D}(\rho)} \mathcal{SN} (|\psi_i\rangle),
\end{align}
where $\mathcal{D}(\rho) = \left\{ p_i, |\psi_i \rangle \right\}$ runs over all decompositions $\rho = \sum_i p_i |\psi_i \rangle \langle \psi_i|$. 

The Schmidt number not only acts as an entanglement certification, but also quantifies the entanglement dimensions. Since computing Eq.~\eqref{eq:DefSNmix} is a demanding task, experimentally practical witnesses for the Schmidt number have been proposed \cite{MalikHighDimensional2026}.  
Here we use the witness based on the trace norm of the partial transpose~\cite{EltschkaPRL2013}, 
\begin{align}\label{eq:WitnessE}
E = \left\vert\left\vert \rho^{T_A} \right\vert\right\vert_1,
\end{align}
where $\rho^{T_A}$ represents the partial transpose with respect to mode A. 
This quantity is directly related to the negativity and logarithmic negativity~\cite{Vidal2002}, and it certifies an entanglement dimension of at least $\lceil E (\rho) \rceil$.

We now adapt this idea to non-Gaussian entanglement certification. We propose the witness 
\begin{align}\label{eq:WitnessNGE}
E_{\text{NG}} := \inf_{\hat U_{G} \in \mathcal{G}_2 } \left\vert\left\vert \left(\hat U_{G} \rho \hat{U}_{\text{G}} ^\dagger \right)^{T_A} \right\vert\right\vert_1.
\end{align}
This quantity evaluates the minimum entanglement that cannot be removed by any Gaussian unitaries $\hat U_G \in \mathcal{G}_2$. Consequently, $E_{\text{NG}}>1$ implies $\rho \notin \mathcal{GE}$ and thus certifies non-Gaussian entanglement. This condition is sharp for pure states: $E_{\text{NG}} (|\psi\rangle) =1$ if and only if the state is Gaussian-entanglable, so that $E_{\text{NG}} (|\psi\rangle) > 1$ certifies genuine non-Gaussian entanglement. As shown in SM~\cite{SM} Sec. I. A, the witness holds $E_{\text{NG}} (\rho) \geq 1$ for all physical states. It satisfies $E_{\text{NG}} (\rho) = 1$ for $\rho \in \mathcal{GE}$, including all separable states and Gaussian states, and is invariant under any Gaussian unitaries $\hat{U}_{\text{G}} \in \mathcal{G}_2$. 
For numerical calculation, we show in SM~\cite{SM} Sec. I. B that the optimization can be restricted to an effective set of Gaussian unitaries $\mathcal{G}^{\text{eff}}_2 \subseteq \mathcal{G}_2$.

\vspace{2mm}
\textbf{A hierarchy of non-Gaussian entanglement.--}
The witness $E_{\rm NG}$ naturally induces a hierarchy through the positive integer
\begin{align}
d := \lceil E_{\text{NG}} (\rho)  \rceil.
\end{align}
The integer $d$ provides a witness on the \textit{non-Gaussian entanglement dimension}, namely the Schmidt number that remains after optimizing over Gaussian unitary transformations. The first nontrivial level $d>1$ certifies non-Gaussian entanglement, while larger values witness higher-dimensional entanglement that cannot be removed by Gaussian unitaries.

For pure states, this hierarchy shows a direct implication for state learning. 
We assume that the Gaussian unitary achieving, or sufficiently approximating, the infimum in Eq.~\eqref{eq:WitnessNGE} is known, and define the corresponding core state $|\psi_{\rm core}\rangle=\hat U_{\rm G(opt)}|\psi\rangle$. Then $d$ is a Schmidt rank witness for $|\psi_{\rm core}\rangle$. Since the trace norm is invariant under unitary transformations, learning the target state $|\psi\rangle$ can be mapped to learning its core $|\psi_{\rm core}\rangle$ followed by the inverse Gaussian unitary. For a finite-energy truncation $|\psi_{\rm core}\rangle \in \mathcal{H}_A^{(D)} \otimes \mathcal{H}_B^{(D)}$ with local dimension $D$, we show in SM~\cite{SM} Sec. I. C that the number of real parameters required to learn the target state $|\psi\rangle$ is lower bounded by
\begin{align}
\mathcal{N}=2d(2D-d)-2.
\end{align}
This lower bound increases monotonically with $d$ for $1\le d\le D$. 
For $d=1$, the state $|\psi\rangle$ is GE and can be mapped to a product core by the optimal Gaussian unitary~\cite{Zhao2025NatComm}. For $d=D$, the bound reaches the full pure-state tomography complexity, $\mathcal{N}=2D^2-2$.

\vspace{2mm}

\textbf{Paradigmatic examples.--}
We now benchmark the criteria $E_{\text{NG}}$ in Eq.~\eqref{eq:WitnessNGE} on representative non-Gaussian states and compare it with the typical entanglement witness $E$ in Eq.~\eqref{eq:WitnessE}. We demonstrate that the resulting integer $d$ witnesses a hierarchy of NGE states. Paradigmatic GE examples with $E_{\text{NG}}=1$, including two-mode cat states, mixtures of two-mode squeezed states and photon-subtractions on pure Gaussian states, are discussed in SM~\cite{SM} Sec. I. D.

\begin{figure}[t]
    \centering
    \includegraphics[width=.39\textwidth]{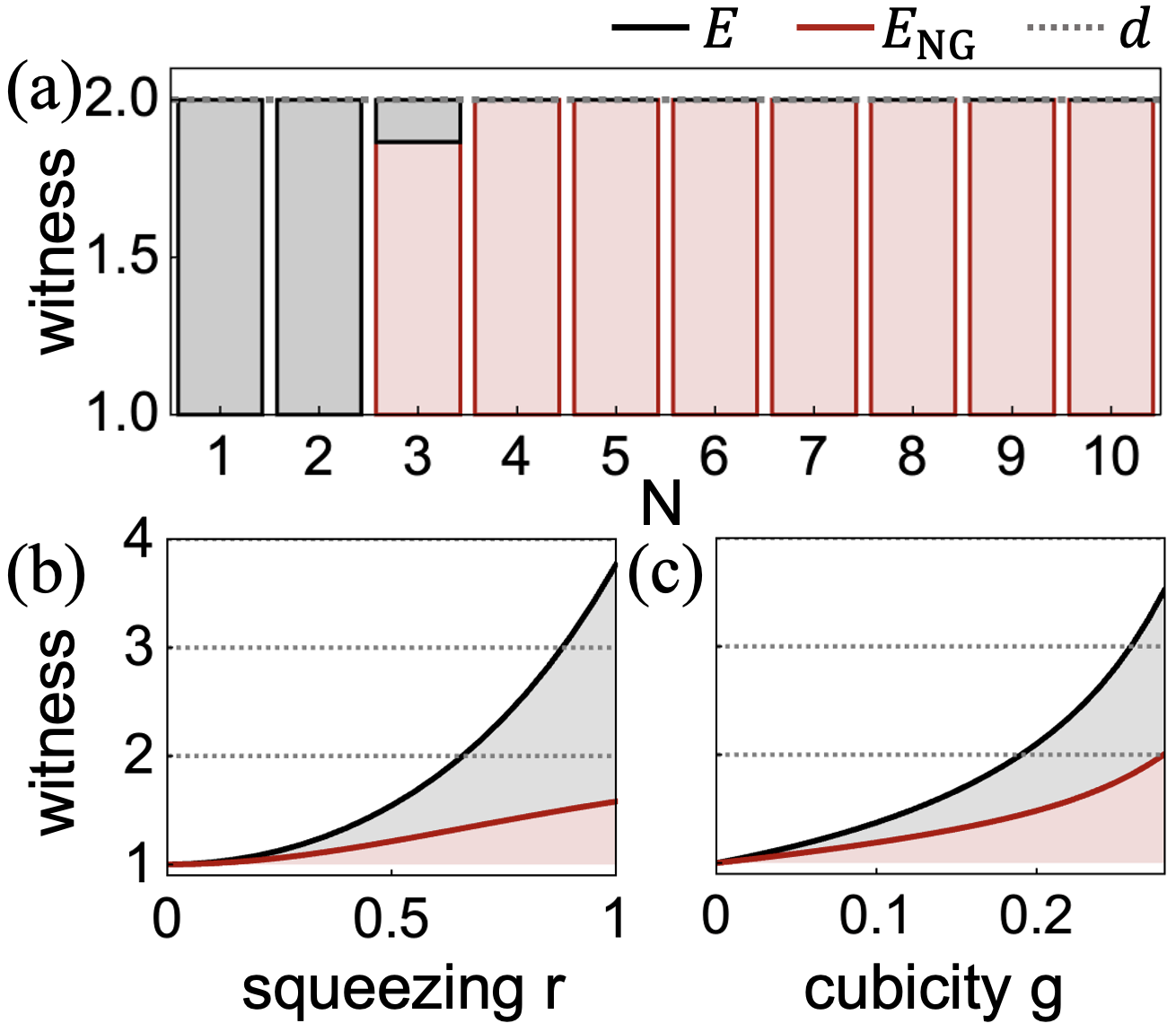}
    \caption{ \textbf{Certifying non-Gaussian entanglement in bipartite pure states. } We compare the typical entanglement witness $E$ in Eq.~\eqref{eq:WitnessE} and the NGE witness $E_{\rm NG}$ in Eq.~\eqref{eq:WitnessNGE} for (a) NOON states $\ket{\psi_{\rm NOON}}$, (b) superpositions of two-mode squeezed-vacuum (TMSV) states $\ket{\psi_{\rm sTMSV}}$, (c) tri-squeezed states $\ket{\psi_{\rm 3-SPDC}}$. Dotted lines represent the non-Gaussian entanglement dimensions witnessed by the integer $d=\lceil E_{\rm NG} \rceil$.  }
    \label{Fig2_NGEpure}
\end{figure}

\vspace{1mm}
\textit{Pure states.--} 
We begin by investigating pure bipartite states beyond the GE class. 
A paradigmatic example is the NOON state,
\begin{align}\label{eq:N00N}
|\psi_{\text{NOON}}\rangle = \frac{1}{\sqrt{2}} \left( |N0 \rangle + |0N\rangle \right).
\end{align}
Although this state has Schmidt rank two for any $N>0$, its non-Gaussian entanglement strongly depends on $N$. As shown in Fig.~\ref{Fig2_NGEpure}(a), the states with $N=\{1,2\}$ are Gaussian-entanglable, whereas $E_{\rm NG}>1$ for $N\geq 3$, certifying genuine non-Gaussian entanglement. This result is consistent with the entanglement-entropy criterion of Ref.~\cite{Zhao2025NatComm}, and shows that higher-order NOON states ($N\geq 3$) cannot be generated by a Hong-Ou-Mandel-type protocol.

Fig.~\ref{Fig2_NGEpure}(b) and (c) present the entanglement for a superposition of two-mode squeezed vacuum (TMSV) states, 
\begin{align}
|\psi_{\text{sTMSV}} \rangle = c \left( \hat{S}_{AB} (r) + \hat{S}_{AB}(-r) \right) |0,0\rangle,
\end{align}
and for states generated by a three-photon spontaneous parametric
down-conversion (SPDC) process~\cite{ChangPRX2020}, 
\begin{align}
|\psi_{\text{3-SPDC}}\rangle = e^{g \left( \hat{a}^\dagger \hat{b}^{\dagger 2} - \hat{a} \hat{b}^2 \right) } |0,0\rangle,
\end{align}
respectively. Here,  $c$ is a normalization constant, $r$ denotes the squeezing strength in two-mode squeezing operators $\hat{S}_{AB}(r) = \exp{\left[ r\left( \hat{a}^\dagger \hat{b}^\dagger - \hat{a} \hat{b} \right) \right]} $, and $g$ represents the cubicity coupling strength. In both cases, entanglement witnesses $E$ and $E_{\text{NG}}$ increase with the nonlinear strength $r$ and $g$.

\begin{figure}[t]
    \centering
    \includegraphics[width=.48\textwidth]{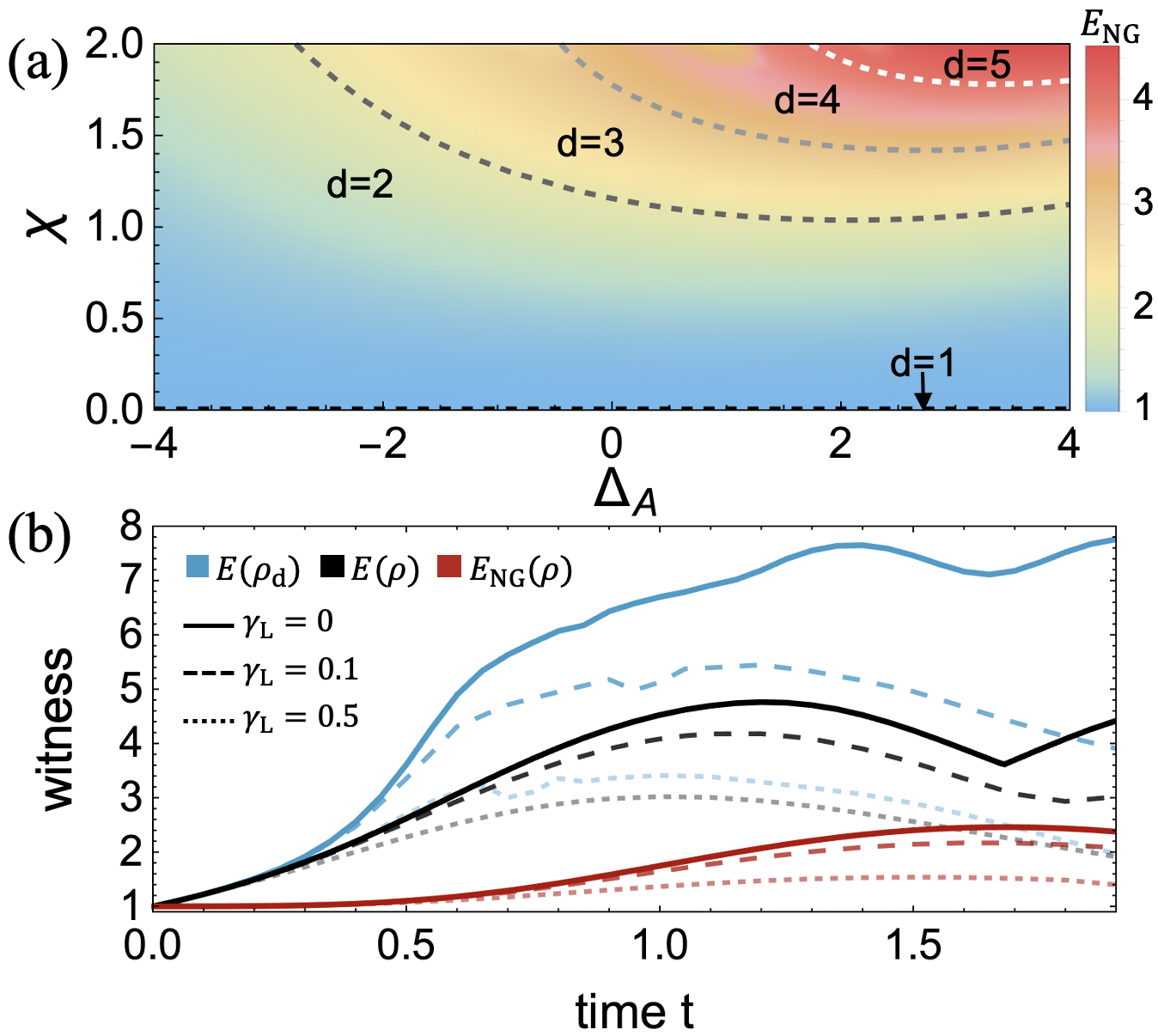}
    \caption{ \textbf{Non-Gaussian entanglement and entanglement distillation in two-mode squeezed Kerr states. } (a) For $t=1, K_A=1$, the NGE witness $E_{\rm NG}$ for pure squeezed-Kerr states changes with $\chi$ and $\Delta$. Dotted contours indicate the witnessed non-Gaussian entanglement dimension $d=\{1,\cdots,5\}$. (b) For $\Delta_A=0, K_A=1,\chi=1$ and different loss rates $\gamma_{\text{L}}=\{0,0.1,0.5\}$, we compare the entanglement witnesses $E(\rho)$ and $E_{\rm NG} (\rho)$ and the distilled entanglement $E(\rho_{\rm d})$ as functions of times $t$.   }
    \label{Fig3_TMSqKerr}
\end{figure}

Then we investigate the non-Gaussian entanglement generated by Kerr nonlinearity.
Kerr nonlinearity is central to a broad range of physics phenomena~\cite{MilburnPRA91,WielingaPRA93,MiranowiczQO1990,ImamogluPRL1997,GuoPRA2024}, and can be conveniently explored in circuit-QED and -QAD platforms~\cite{GerhardNature2013,GrimmNature2020,marti_quantum_2024}. 
Here, we consider a bipartite bosonic system described by the Hamiltonian,
\begin{equation}\label{eq:Heff}
    \hat{H} = \Delta_A \hat{a}^\dagger \hat{a} + \chi \hat{a}^\dagger \hat{b}^\dagger + \chi^* \hat{a} \hat{b} - K_A \hat{a}^{\dagger 2} \hat{a}^2 .
\end{equation}
Here, $\Delta_A$ is a local detuning, $\chi$ is the two-mode squeezing strength, and $K_A$ is a self-Kerr nonlinearity acting on mode $A$. Starting from the two-mode vacuum, the system evolves as $|\psi_{\text{sq-Kerr}}\rangle = e^{-i\hat{H}t}|0,0\rangle$. For $K_A=0$, the evolution is Gaussian. When Kerr nonlinearity is included, the quartic interaction drives the state into the non-Gaussian regime.
Fig.~\ref{Fig3_TMSqKerr}(a) shows $E_{\text{NG}}$ for $K_A=1$ and $t=1$. For $\chi=0$, the evolution is local and no entanglement is generated. For $\chi>0$, $E_{\text{NG}}$ grows with the squeezing strength $\chi$ and strongly depends on the detuning $\Delta_A$. 
Moreover, the dotted contours mark regions with different $d$, revealing a hierarchy of non-Gaussian entanglement generated by the interplay between a self-Kerr nonlinearity and a two-mode squeezing interaction. 
It indicates that non-Gaussian entanglement dimension grows with the detuning $\Delta_A$ and squeezing $\chi$, and that the complexity of learning the states increases accordingly.

\vspace{1mm}
\textit{Mixed states.--}
In realistic scenarios, inevitable experimental imperfections result in mixed states, thus motivating us to examine the robustness of our criteria in the presence of noise.

The Hamiltonian Eq.~\eqref{eq:Heff} can be realized in state-of-the-art circuit QED and QAD platforms~\cite{WangScience2016,GaoNature2019,ChuNP2024}, where energy relaxation can be considered as one of the major noise sources. 
This can be conveniently analyzed by a loss model, where jump operators, $\sqrt{\gamma_{A}} \hat{a}$ and $\sqrt{\gamma_{B}} \hat{b}$, are introduced to the Lindblad master equations. 
To simplify the model, we take equal loss rates $\gamma_A = \gamma_B = \gamma_{\text{L}}$.
Fig.~\ref{Fig3_TMSqKerr}(b) shows $E(\rho)$ (black) and $E_{\text{NG}}(\rho)$ (red) as functions of the evolution time for different loss rates. As time accumulates, $E$ and $E_{\text{NG}}$ increase and then decrease, yielding optimal evolution times. The maximum of $E_{\text{NG}}$ appears at a later time than the one from $E$. As expected, loss reduces both entanglement witnesses.
For $\gamma_{\text{L}} = 0.1$,  $88\%\, E$ and $89\%\, E_{\text{NG}}$ are retained compared to the lossless case. For $\gamma_{\text{L}} = 0.5$, the retained values are about $63\%$ for both quantities.

\vspace{2mm}
\textit{Application to entanglement distillation.--} 
It is known that Gaussian entangled states cannot be distilled in the Gaussian settings~\cite{EisertPRL2002}. 
We show that the NGE states detected by $E_{\rm NG}>1$ provide a promising resource in entanglement distillation. 
We consider a two-copy distillation protocol as illustrated in SM~\cite{SM} Sec. II, and use the lossy two-mode squeezed Kerr states $\rho$ as the input. With appropriate choices of Gaussian unitaries and homodyne measurements, the entanglement distillation can be successfully achieved, where the final state $\rho_{\rm d}$ exhibits enhanced entanglement than the input, $E(\rho_{\rm d})>E(\rho)$. 
In Fig.~\ref{Fig3_TMSqKerr}, we plot the distilled entanglement $E(\rho_{\rm d})$ (blue) as a function of time, and observe $E(\rho_{\rm d})> E(\rho)$ within the whole time period considered. 
For short times ($t<0.4$), the non-Gaussian entanglement $E_{\rm NG}(\rho)$ is quite small, so that the corresponding distillation gain $E(\rho_{\rm d})$ is low. For larger times, a significant enhancement is shown when the non-Gaussian entanglement becomes more pronounced.

\vspace{2mm}
\textbf{Experimentally practical NOON-type GNGE witness.--}
So far, our criterion in Eq.~\eqref{eq:WitnessNGE} provides a general characterization of non-Gaussian entanglement, but it is based on the density matrix, thus requires full state tomography, which is particularly costly in CV platforms. We now show that, for target states in a simple form, such as NOON states, genuine non-Gaussian entanglement can be certified from detecting only a few density-matrix elements.

For a given state $\rho$, we construct the NOON-type witness $f=\text{Tr}[\mathcal{F}\rho]$, where
\begin{align}\label{eq:N00Nwitness}
\mathcal{F} &= \frac{1}{2} \left( |0,N \rangle \langle 0, N| + |N,0 \rangle \langle N,0| + |0,N \rangle \langle N, 0| +|N,0 \rangle \langle 0,N| \right).
\end{align}
The maximum value achieved by any pure GE states is
\begin{align}
f_{\text{G}} := \max_{\hat{U}_{\text{G}} \in \mathcal{G}_2, |\psi_A\rangle, |\psi_B \rangle } \langle \psi_A | \otimes \langle \psi_B | \hat{U}_{\text{G}}^\dagger \mathcal{F} \hat{U}_{\text{G}} |\psi_A \rangle \otimes |\psi_B \rangle.
\end{align}
As shown in SM~\cite{SM} Sec. III. A, this GE threshold can be expressed by, 
\begin{align}
f_{\text{G}} = \max_{ \hat{U}_{\text{G}} \in \mathcal{G}_2 } \lambda_{\max}^2 \left( \hat{U}_{\text{G}} |\psi_{\text{NOON}}  \rangle \right),
\end{align}
where $\lambda_{\max} \left( |\psi\rangle \right)$ is the maximum Schmidt coefficient for $|\psi\rangle$, which is a central quantifier of the optimal separable approximation to a bipartite state, thereby providing a natural threshold for detecting and characterizing entanglement~\cite{BourennanePRL2004}. Analogously, the optimization can be restricted to an effective family $\mathcal{G}^{\text{eff}}_2 \subseteq \mathcal{G}_2$.
Moreover, this bound remains within the range  (see SM~\cite{SM} Sec. III. B)
\begin{align}\label{eq:NOONbounds}
\max \left\{ 2^{-1}, 2^{1-N} \binom{N }{\lfloor N/2 \rfloor } \right\}   \leq f_{\text{G}} \leq 1.
\end{align}
Although $f_{\rm G}$ is derived from pure GE states, we prove in SM~\cite{SM} Sec. III. C that the same threshold applies to any state $\rho \in \mathcal{GE}_{\text{mix}}$. As a result, any state satisfying $f>f_{\text{G}}$ is certified as genuine non-Gaussian entanglement. 

\begin{figure}[t]
    \centering
    \includegraphics[width=.48\textwidth]{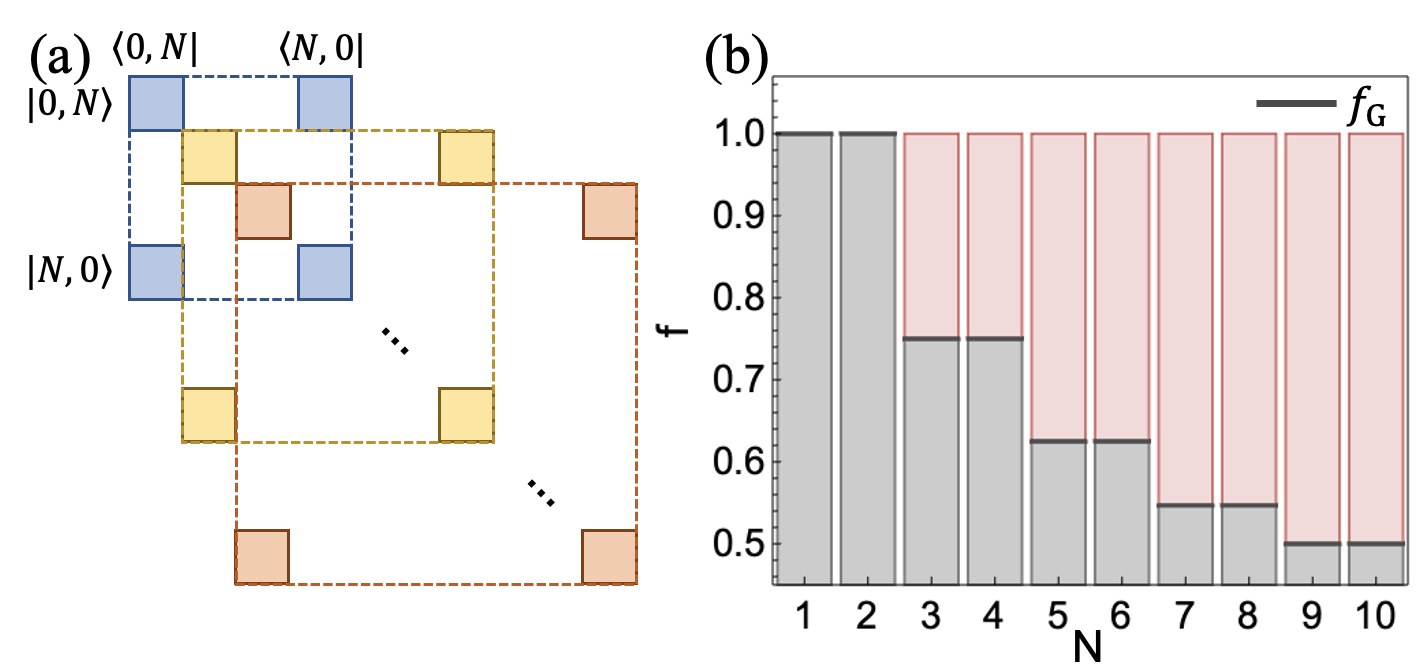}
    \caption{ \textbf{Experimentally practical NOON-type witness for genuine non-Gaussian entanglement. } (a) The witness requires only four density-matrix elements in the subspace of target NOON states. (b) The Gaussian-entanglable threshold $f_{\text{G}}$ for different $N$. If the detection statistics $f$ lies in the red regions, $f> f_{\text{G}}$, certifies genuine non-Gaussian entanglement. 
    }
    \label{Fig4_NGEwitness}
\end{figure}

Fig.~\ref{Fig4_NGEwitness} shows the threshold $f_{\text{G}}$ for different NOON states. If we have a priori knowledge of the target NOON states to be certified, only four density-matrix elements specified by Eq.~\eqref{eq:N00Nwitness} are required, which is much more experimentally practical than full tomography. For $N=\{1,2\}$, we find $f_{\rm G}=1$, so that no violation is possible, which agrees with the fact in Fig~\ref{Fig2_NGEpure}(a) that these NOON states are GE states. For $3 \leq N \leq 10$, the threshold $f_{\text{G}}$ decreases with $N$. We numerically obtain $f_{\text{G}}=2^{1-N}\binom{N}{\lfloor N/2 \rfloor}$ for $3 \leq N \leq 8$, and $f_{\text{G}}=1/2$ for $N=\{9,10\}$, coinciding with the lower bounds in Eq.~\eqref{eq:NOONbounds}.
Remarkably, within the range considered here, $N \leq 10$, the decreasing threshold $f_{\rm G}$ implies that it is easier to certify genuine non-Gaussian entanglement in higher-$N$ NOON states.

\vspace{2mm}
\textbf{Conclusions.--}
In this work, we develop  witnesses for the certification and characterization of Gaussian-entanglable (GE), non-Gaussian entangled (NGE), and genuinely non-Gaussian entangled (GNGE) states in bipartite continuous-variable systems.
Our main witness of certifying NGE is based on the partial transpose of the density matrix, minimized over all two-mode Gaussian unitaries. 
This quantity provides a lower bound on the Schmidt number irreducible by Gaussian transformations, thus defines a natural hierarchy of witnessing non-Gaussian entanglement dimensions. 
For pure states, the condition becomes sufficient and necessary for certifying GNGE states, and the hierarchy is connected to the complexity of the state learning.
We have benchmarked the framework on paradigmatic non-Gaussian states, and examined its robustness against loss.
To simplify the experimental implementation, we further complemented our criteria with a practical witness tailored to NOON states, where only four density-matrix elements are required for detection.

Our results extend the Schmidt-number paradigm to the CV non-Gaussian entanglement regime and provide a practical route to identifying entanglement resources. 
As an example, we show that the certified non-Gaussian entanglement in squeezed-Kerr states is an effective resource for entanglement distillation, highlighting promising applications of witnessed non-Gaussian entanglement.
More broadly, non-Gaussian entangled states are emerging candidates in various quantum information fields, such as quantum metrology, quantum computing and quantum communication. 
Our work opens new avenues for entanglement characterization and for identifying resources that are essential for quantum technologies.

\textit{Note added.--}%
Recently, we became aware of related independent work on fidelity-based certification of genuine non-Gaussian entanglement~\cite{MattiaArXiv2026}.

\vspace{1mm}
\textit{Acknowledgments.--}%
This work is supported by Quantum Science and Technology-National Science and Technology Major Project (Grant No. 2024ZD0302401 and No. 2021ZD0301500), National Natural Science Foundation of China (No. 12125402, No. 12534016, No. 12505010, No. 12447157, No. 12405005), and Beijing Natural Science Foundation (Grant No. Z240007). J.G. acknowledges Postdoctoral Fellowship Program of CPSF (GZB20240027), and the China Postdoctoral Science Foundation (No. 2024M760072). S.L. acknowledges the China Postdoctoral Science Foundation (No. 2023M740119). 
M.F. was supported by the Swiss National Science Foundation Ambizione Grant No. 208886, and by The Branco Weiss Fellowship -- Society in Science, administered by the ETH Z\"{u}rich.

\bibliography{mybib}

\clearpage
\newpage

\begin{widetext}

\section{Supplemental material for ``Non-Gaussian Entanglement Hierarchy Based on the Schmidt Number''}

\section{I.\quad Non-Gaussian Entanglement Witness}

\subsection{A.\quad Properties of non-Gaussian entanglement witness $E_{\text{NG}}$}

In this section, we prove some properties of the non-Gaussian entanglement witness,
\begin{align}\label{SMeq:WitnessNGE}
E_{\text{NG}} := \inf_{\hat U_{G} \in \mathcal{G}_2 } \left\vert\left\vert \left(\hat U_{G} \rho \hat{U}_{\text{G}} ^\dagger \right)^{T_A} \right\vert\right\vert_1,
\end{align}
where $\hat{U}_{\rm G}\in\mathcal{G}_2$ is a two-mode Gaussian unitary, and $\rho^{T_A}$ is the partial transpose with respect to mode $A$.

\subsubsection{a.\quad $E_{\text{NG}} \geq 1$ for all physical states}

For any physical state $\rho$ and any Gaussian unitary $\hat{U}_{\text{G}}$, the resulting state $\rho_{\text{G}} = \hat{U}_{\text{G}} \rho \hat{U}_{\text{G}}^\dagger$ satisfies $\rho_{\text{G}} \geq 0$ and $ \text{Tr}\left( \rho_{\text{G}} \right) =1$. Its partial transpose operator $\rho_{\text{G}}^{T_A}$ is Hermitian and has unit trace $\operatorname{Tr} \left( \rho_{\text{G}}^{T_A} \right)=1$. We denote the eigenvalues of $\rho_{\text{G}}^{T_A}$ as $\{\mu_i\}$, then the trace is $\text{Tr}\left( \rho_{\text{G}}^{T_A} \right)=\sum_i \mu_i=1$ and the trace norm is $ \left\vert\left\vert \rho_{\text{G}}^{T_A} \right\vert\right\vert_1 = \sum_i |\mu_i| $. Using $\sum_i |\mu_i| \geq \left\vert \sum_i \mu_i \right \vert $, we obtain
\begin{align}
\left\vert\left\vert \rho_{\text{G}}^{T_A} \right\vert\right\vert_1 \geq \left\vert\operatorname{Tr} \left( \rho_{\text{G}}^{T_A} \right) \right\vert = 1,
\end{align}
which implies
\begin{align}
E_{\rm NG}(\rho)\geq 1
\end{align}
for arbitrary physical states $\rho$.

\subsubsection{b.\quad $E_{\text{NG}} = 1$ for all separable states}

For a separable state $\rho_{\rm sep}=\sum_i c_i\rho_A^{(i)}\otimes\rho_B^{(i)}$, with $c_i \geq 0, \sum_i c_i = 1$, its partial transpose on mode $A$ is 
\begin{align}
\rho_{\rm sep}^{T_A} = \sum_i c_i \left( \rho_A^{(i)} \right)^T \otimes \rho_B^{(i)}.
\end{align}
Since $\rho_{\rm sep}^{T_A}$ is positive and has unit trace, we have
\begin{align}
\left\vert\left\vert \rho_{\rm sep}^{T_A} \right\vert\right\vert_1 = \text{Tr} \left( \rho_{\rm sep}^{T_A} \right)=1.
\end{align}
If we choose $\hat{U}_{\text{G}} = \hat{I}$ in Eq.~\eqref{SMeq:WitnessNGE}, we have $E_{\rm NG}(\rho_{\rm sep}) \leq 1$. Considering the conditions $E_{\text{NG}}(\rho)\geq 1$,  we thus obtain
\begin{align}
E_{\rm NG}(\rho_{\rm sep})=1.
\end{align}

\subsubsection{c.\quad $E_{\text{NG}} = 1$ for all Gaussian states}

By the Williamson decomposition~\cite{Weedbrook2012}, any two-mode Gaussian state can be written as
\begin{align}
\rho_{\text{GS}} = \hat{V}_{\text{G}} \left[ \rho_{\rm th}(n_1)\otimes\rho_{\rm th}(n_2) \right] \hat{V}_{\text{G}}^{\dagger} ,
\end{align}
where $\rho_{\rm th}(n_i)$ are single-mode thermal states with local particle number $n_i$ on mode $i$, and $\hat{V}_{\text{G}}$ denotes a two-mode Gaussian unitary. 
Choosing $\hat{U}_{\text{G}}=\hat{V}_{\text{G}}^{\dagger}$ in Eq.~\eqref{SMeq:WitnessNGE}, one obtains
\begin{align}
\hat U_{\text{G}} \rho_{\rm GS} \hat U_{\text{G}}^{\dagger}
=
\rho_{\rm th}(n_1)\otimes\rho_{\rm th}(n_2),
\end{align}
which is a product of thermal states. Using $\left\vert\left\vert \rho_{\rm sep}^{T_A} \right\vert\right\vert_1 = 1$, we have
\begin{align}
\left \vert \left \vert
\left[ \rho_{\rm th}(n_1)\otimes\rho_{\rm th}(n_2) \right]^{T_A}
\right\vert \right\vert_1 = 1 .
\end{align}
This implies $E_{\rm NG}(\rho_\text{GS})\leq 1 $. Together with the condition $E_{\rm NG}(\rho) \geq 1$, we prove that
\begin{align}
E_{\rm NG}(\rho_\text{\text{GS}})=1
\end{align}
for arbitrary two-mode Gaussian state.

\subsubsection{d.\quad $E_{\text{NG}} = 1$ for all Gaussian-entanglable states}
By definition, any $\rho_{\rm GE}\in\mathcal{GE}$ can be written as $\rho_{\rm GE}=\hat{V}_{\rm G}\rho_{\rm sep}\hat {V}_{\rm G}^\dagger$, where $\rho_{\rm sep}$ is a separable state and $\hat{V}\in\mathcal{G}_2$ is a two-mode Gaussian unitary.
Choosing $\hat U_{\rm G}=\hat V_{\rm G}^\dagger$ in Eq.~\eqref{SMeq:WitnessNGE}, we have $\hat{U}_{\rm G}\rho_{\rm GE}\hat{U}_{\rm G}^\dagger = \rho_{\rm sep}$.
Using $\left \| (\rho_{\rm sep})^{T_A} \right\|_1=1$, we obtain
\begin{align}
E_{\rm NG} (\rho_{\rm GE}) =1,
\end{align}
for all $\rho_{\rm GE}\in\mathcal{GE}$.

\subsubsection{e.\quad $E_{\rm NG} = 1$ is a sufficient and necessary condition for pure GE states}

After proving that all GE states satisfy $E_{\rm NG} = 1$, we now show that this criterion is both sufficient and necessary condition for pure states.

For a two-mode pure state $\ket{\psi}$ satisfying $E_{\rm NG} (\ket{\psi} ) = 1$, we assume that the infimum in the criterion is saturated by an optimal Gaussian unitary $\hat{U}_{\rm G}^{\rm opt}$, and define $\ket{\phi} = \hat{U}_{\rm G}^{\rm opt} \ket{\psi}$. By the Schmidt decomposition $\ket{\phi} = \sum_i \lambda_i |n_i \rangle |m_i\rangle$ with $\lambda_i \geq 0 , \sum_i \lambda_i^2 =1$, it gives
\begin{align}
\left \vert \left \vert \left( \ket{\phi} \bra{\phi}  \right)^{T_A} \right \vert \right \vert_1 = \left( \sum_i \lambda_i \right)^2 = 1.
\end{align}
Since $\sum_i \lambda_i^2 = \left( \sum_i \lambda_i \right)^2 =1 $ and $\lambda_i \geq 0$, these equalities hold if and only if only one Schmidt coefficient is positive. This indicates $\ket{\phi}$ is a product state, and therefore $\ket{\psi} = (\hat{U}_{\rm G}^{\rm opt})^{\dagger} \ket{\phi} $ is a pure GE state. 
This derivation is based on the fact that, for bipartite pure states, the Peres-Horodecki positive partial transpose (PPT) criterion is a sufficient and necessary condition, which implying that the trace norm of partial-transpose equals one if and only if the pure state is a product state~\cite{Vidal2002,Peres1996,Horodecki1997}.

Consequently, if the infimum in $E_{\rm NG}$ is attained, the criterion provides a necessary and sufficient condition for pure states, i.e., $E_{\rm NG}(|\psi\rangle) = 1$ if and only if $\ket{\psi}$ is a pure GE state, and $E_{\rm NG}(|\psi\rangle) > 1$ certifies genuine non-Gaussian entanglement.

\subsection{B.\quad The effective two-mode Gaussian unitary}

The optimization over two-mode Gaussian unitaries can be simplified~\cite{Zhao2025NatComm}, by removing elements that do not affect the witness in Eq.~\eqref{SMeq:WitnessNGE}. A general two-mode Gaussian unitary can be expressed by
\begin{align}
\hat{U}_{\text{G}} = \left(  \hat{D}_A(\alpha_1) \otimes \hat{D}_B(\alpha_2) \right) \hat{V}_{\text{II}} \left( \hat{S}_A(r_1) \otimes \hat{S}_B(r_2) \right) \hat{V}_{\text{I}},
\end{align}
where $\hat{D}(\alpha_i) = e^{\alpha_i \hat{a}_i^\dagger -\alpha_i^* \hat{a}_i } $ and $\hat{S}(r_i) = e^{r_i/2(\hat{a}_i^2 - \hat{a}_i^{\dagger 2})}$ are single-mode displacement and squeezing operators on mode $i$, respectively. $\hat{V} = \hat{R}_A(\theta_A) \otimes \hat{R}_B(\theta_B) \hat{U}_{\text{BS}} (\theta) \hat{R}_A (\theta_0) $ denotes the passive interferometer, which involves the local rotation operators $\hat{R}(\theta_i) = e^{-i \theta_i \hat{a}^\dagger \hat{a}}$ and the beam-splitter operation $\hat{U}_{\text{BS}} =e^{\theta \left( \hat{a}^\dagger \hat{b} - \hat{a}\hat{b}^\dagger \right)} $.
Therefore, an arbitrary two-mode Gaussian unitary contain $14$ real variables:
\begin{align}
\hat{U}_{\text{G}} &= \left( \hat{D}_A(\text{Re}(\alpha_1), \text{Im}(\alpha_1) ) \otimes \hat{D}_B( (\text{Re}(\alpha_2), \text{Im}(\alpha_2) ) ) \right) \left( \hat{R}_A(\theta^{\text{II}}_1) \otimes \hat{R}_B (\theta^{\text{II}}_2)  \right) \nonumber \\
&\; \cdot \hat{U}_{\text{BS}} (\theta^{\text{II}}) \hat{R}_A (\theta^{\text{II}}_0) \left( \hat{S}_A (r_1) \otimes \hat{S}_B (r_2) \right) \left( \hat{R}_A(\theta^{\text{I}}_1) \otimes \hat{R}_B (\theta^{\text{I}}_2)  \right) \hat{U}_{\text{BS}} (\theta^{\text{I}}) \hat{R}_A (\theta^{\text{I}}_0).
\end{align}
Since $\left\| \left( \hat{U}_{\text{G}} \rho \hat{U}_{\text{G}}^\dagger \right)^{T_A} \right\|_1$ is invariant under local Gaussian unitaries applied after the transformation, the $\hat{D}_A(\alpha_1) \otimes \hat{D}_B(\alpha_2)$ and $\hat{R}_A(\theta^{\text{II}}_1) \otimes \hat{R}_B (\theta^{\text{II}}_2)$ can be discarded. This removes 6 parameters.
The remaining Gaussian unitary can be expressed as
\begin{align}
\hat{U}_{\text{G(8)}} &= \hat{U}_{\rm BS}(\theta^{\rm II}) \hat{R}_A(\theta^{\rm II}_0) \left[ \hat{S}_A(r_1)\otimes \hat{S}_B(r_2) \right] \hat{V}_{\text{I}} \nonumber \\
&= \hat{U}_{\rm BS}(\theta^{\rm II}) \left[ \hat{S}_A (r_2) \otimes \hat{S}_B (r_2) \right] \left[ \hat{S}_A (-r_2) \hat{R}_A(\theta^{\rm II}_0)  \hat{S}_A(r_1)\otimes \hat{I}_B \right] \hat{V}_{\text{I}} \nonumber \\
&= \left[ \hat{S}_A (r_2) \otimes \hat{S}_B (r_2) \right] \hat{U}_{\rm BS}(\theta^{\rm II}) \left[ \hat{S}_A (-r_2) \hat{R}_A(\theta^{\rm II}_0)  \hat{S}_A(r_1)\otimes \hat{I}_B \right] \hat{V}_{\text{I}}
\end{align}
Here we use the commutation relation, $\left[\hat{S}_A (r_2) \otimes \hat{S}_B (r_2), \hat{U}_{\rm BS}(\theta^{\rm II}) \right] =0 $. Then two local unitaries $\hat{S}_A (r_2) \otimes \hat{S}_B (r_2)$ can be discarded. By Bloch-Messiah decomposition, we can reexpress $\hat{S}_A (-r_2) \hat{R}_A(\theta^{\rm II}_0)  \hat{S}_A(r_1) = \hat{R}_A (\phi_1) \hat{S}_A(r) \hat{R}_A(\phi_2)$. After we relabel $\phi_1 \rightarrow \theta^{\text{II}}_0$ and $\phi_2+\theta^{\rm I}_1 \rightarrow \theta^{\rm I}_1$, we obtain the effective Gaussian unitary containing seven real parameters,
\begin{align}\label{SMeq:UGeff}
\hat{U}_{\text{G}}^{\rm eff} 
&= \hat{U}_{\text{BS}} (\theta^\text{II}) \left[ \hat{R}_A(\theta^{\text{II}}_0) \hat{S}_A(r) \hat{R}_A(\theta^{\text{I}}_1) \otimes  \hat{R}_B (\theta^{\text{I}}_2) \right] \hat{U}_{\text{BS}} (\theta^{\text{I}}) \hat{R}_A (\theta^{\text{I}}_0).
\end{align}
We denote the set of all such unitaries by $\mathcal{G}_2^{\text{eff}} \subseteq \mathcal{G}_2$. In this way, the numerical optimization can be mapped to 
\begin{align}
E_{\rm NG} (\rho) = \inf_{\hat{U}_{\rm G}^{\rm eff} \in \mathcal{G}_2^{\rm eff} } \left \| \left( \hat{U}_{\rm G}^{\rm eff} \rho \hat{U}_{\rm G}^{\rm eff \dagger} \right)^{T_A} \right\|_1.
\end{align}

\subsection{C.\quad Efficient learning of pure non-Gaussian entangled states}

In Ref.~\cite{Zhao2025NatComm}, Gaussian disentangling tomography was proposed as an efficient method to learn Gaussian-entanglable (GE) states. The protocol first estimates the displacement vector and covariance matrix using homodyne and heterodyne detection. These first-and second-order moments identify a Gaussian unitary, which maps a pure GE state to a passive-separable or separable state. The task is then mapped to local state tomography, thereby reducing the complexity of learning GE states.

Here we adopt this idea for states beyond the GE class. Instead of considering whether the state can be disentangled, we quantify the non-Gaussian entanglement by the quantity $E_{\text{NG}}$, and show that the positive integer $d=\lceil E_{\text{NG}}\rceil$ provides a lower bound on the complexity of learning a pure state.

We consider a two-mode pure state $|\psi \rangle\in \mathcal H_A\otimes\mathcal H_B$, with $ E_{\rm NG}(|\psi\rangle) = \inf_{\hat U_G\in\mathcal G_2} \left\| \left(\hat U_G |\psi\rangle\langle\psi| \hat U_G^\dagger \right)^{T_A} \right\|_1 $ and the positive integer $d = \left\lceil E_{\rm NG}(|\psi\rangle) \right\rceil$.
We assume the infimum can be achieved by a known Gaussian unitary $\hat U_{\rm G(opt)}$, and then define the corresponding core state
\begin{align}
|\psi_{\rm core}\rangle = \hat U_{\rm G(opt)} |\psi \rangle .
\end{align}
We therefore obtain $E_{\text{NG}} (|\psi\rangle) = E_{\text{NG}} (|\psi_{\rm core}\rangle)$. The Schmidt decomposition of the core state is $|\psi_{\rm core}\rangle = \sum_{i=1}^{r} \lambda_i |n_i\rangle |m_i\rangle$, where $\lambda_i>0, \sum_i \lambda_i^2=1$ and $r=\mathcal{SN}(|\psi_{\rm core} \rangle)$ is the Schmidt rank. Using $ E_{\text{NG}} (\psi_{\rm core}) =  \left\| \left( |\psi_{\rm core}\rangle\langle\psi_{\rm core}| \right)^{T_A} \right\|_1 $ and the Cauchy-Schwarz inequality $ \left( \sum_{i=1}^{r}\lambda_i \right)^2 \leq r \sum_{i=1}^{r}\lambda_i^2 = r$, we obtain $r \geq E_{\rm NG} (|\psi \rangle)$ and $r \geq d$, which indicates that $d$ is a Schmidt rank witness for $|\psi_{\rm core}\rangle$.

Since the trace norm is invariant under the unitary transformation, we show that the trace distance between the true state $|\psi\rangle$ and the reconstructed state $|\psi'\rangle$ is equivalent to the trace distance between the true core state $|\psi_{\rm core}\rangle$ and the reconstructed core state, 
\begin{align}
\frac{1}{2} \left\| | \psi'\rangle \langle \psi' | - | \psi\rangle \langle \psi |   \right\|_1 = \frac{1}{2} \left\| | \psi_{\rm core}'\rangle \langle \psi_{\rm core}' | - | \psi_{\rm core}\rangle \langle \psi_{\rm core} |   \right\|_1\leq \epsilon,
\end{align}
where $\epsilon$ is the error. Based on the prior knowledge of $\hat U_{\rm G(opt)}$, we set the estimate $|\psi' \rangle = \hat{U}_{\text{G}(\rm opt)}^\dagger |\psi_{\rm core}'\rangle $. In this way, a tomography of $|\psi\rangle$ can be mapped to a tomography of $|\psi_{\rm core}\rangle$ by applying $\hat U_{\rm G(opt)}$.

We consider a finite-energy truncation and restrict each mode to the local Hilbert space $\mathcal H_{A(B)} = \text{span}\{|0\rangle, \cdots,|D-1\rangle \}$ with local dimensions $D$. After removing the parameters from the normalization and the global phase shift, we obtain the number of real parameters required for a full tomography of a two-mode pure state
\begin{align}
\mathcal{N}_{\text{full}} = 2 D^2 -2.
\end{align}
A pure core state can be reexpressed as $|\psi_{\rm core} \rangle = \sum_{i,j=0}^{D-1} C_{i,j}|\mu_i\rangle |\nu_j \rangle$, where the Schmidt rank $r$ is equal to the rank of the matrix $C$. Independent real parameters to characterize a complex matrix with $D\times D$ dimensions and rank $r$ is $2r (2D -r)$. After considering the normalization and the global phase constraint, we obtain $\mathcal{N}_{\text{core}} (r) = 2r (2D -r) -2$.
Since the non-Gaussian entanglement integer $d$ is a Schmidt number witness, i.e., $d\leq r$, and $\mathcal{N}_{\rm core}(r)$ increases monotonically with $r$ for $1\leq r \leq D$, the number of real parameters required to learn the core state is lower bounded by
\begin{align}
\mathcal{N}_{\text{core}} (d)=2d(2D-d)-2.
\end{align}

In Fig.~\ref{SMFig1_tomography}, we present the $\mathcal{N}_{\text{core}} (d)$ as a function of the non-Gaussian entanglement integer $d$. $\mathcal{N}_{\rm core} (d)$ increases monotonically with $d$. For $d=1$, $|\psi\rangle$ is a GE state, and can be mapped to a product core state by the optimal Gaussian unitary. For $d=D$, the complexity returns to the full tomography $\mathcal{N}_{\text{core}}(d=D) = \mathcal{N}_{\text{full}}$.

\begin{figure}[h]
    \centering
    \includegraphics[width=.48\textwidth]{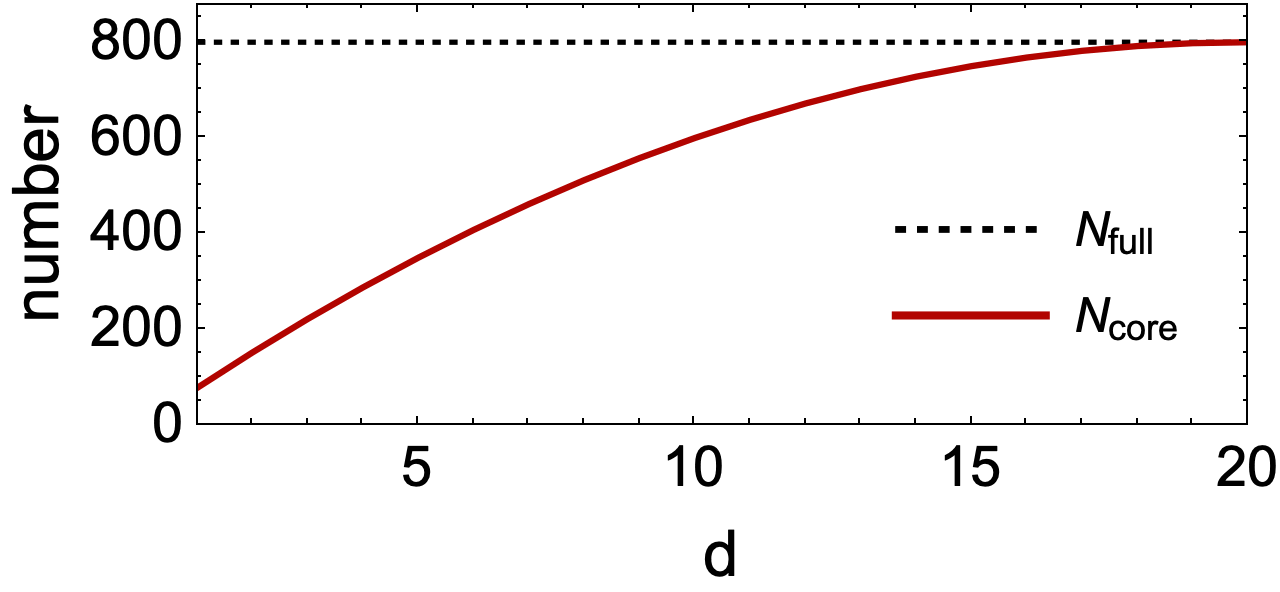}
    \caption{ 
    The complexity of pure state learning in a finite-energy truncation with local dimension $D=20$. The red solid line, $\mathcal N_{\rm core}(d)=2d(2D-d)-2$, provides a lower bound on the number of real parameters required to learn the pure state, while the black dashed line, $\mathcal N_{\rm full}=2D^2-2$, shows the full tomography complexity. The witnessed integer $d=\lceil E_{\rm NG}\rceil$ therefore reflects the complexity of learning two-mode pure states.}
    \label{SMFig1_tomography}
\end{figure}

\subsection{D.\quad More examples for $E_{\text{NG}} = 1$}
\subsubsection{a.\quad Two mode cat states}

We now show that the two-mode pure cat state is a Gaussian-entanglable state. We consider $\alpha \in \mathbb{R}$, and write the two-mode cat state as 
\begin{align}
|\psi_{\text{cat}}\rangle = c \left(|\alpha,\alpha\rangle + |-\alpha,-\alpha\rangle \right),
\end{align}
with the normalization values $c=\left[2\left(1+e^{-4\alpha^2}\right) \right]^{-1/2}$.

If we choose the Gaussian unitary as a balanced beam splitter, $\hat{U}_{\text{G}} = \hat{U}_{\text{BS}}(\pi/4) = \exp{\left[ \pi/4 \left( \hat{a}^\dagger \hat{b} - \hat{a} \hat{b}^\dagger \right) \right]}$, we will have $\hat{U}_{\text{BS}}(\pi/4) |\alpha, \alpha\rangle = |\sqrt{2} \alpha \rangle \otimes |0\rangle $ and $\hat{U}_{\text{BS}}(\pi/4) |-\alpha, -\alpha\rangle = |-\sqrt{2} \alpha \rangle \otimes |0\rangle $. This gives 
\begin{align}
\hat{U}_{\text{BS}} (\pi/4) |\psi_{\text{cat}} \rangle &= c \left( \ket{\sqrt{2}\alpha} + \ket{-\sqrt{2} \alpha} \right)\otimes |0\rangle,
\end{align}
which is a separable state. Using $E_{\text{NG}} (\rho_{\text{sep}})=1$, we obtain
\begin{align}
E_{\text{NG}}\left( |\psi_{\text{cat}}\rangle \right)=1.
\end{align}

\subsubsection{b.\quad Classical mixture of two-mode squeezed vacuum (TMSV) states }

We now consider a classical mixture of two two-mode squeezed vacuum (TMSV) states,
\begin{align}
\rho_{\rm mix(TMSV)} = p|\mathrm{TMSV}(\xi_1)\rangle\langle \mathrm{TMSV}(\xi_1)| + (1-p)|\mathrm{TMSV}(\xi_2)\rangle\langle \mathrm{TMSV}(\xi_2)|,
\end{align}
where $\xi_j=r_j e^{i\theta_j}$ is determined by the squeezing strength $r_j$ and squeezing direction $\theta_j$, and TMSV states are generated by applying the two-mode squeezing operator $\hat{S}_{AB}(\xi)= \exp {\left[\xi \hat a^\dagger \hat b^\dagger - \xi^* \hat a\hat b \right]}$ on the vacuum, $|\text{TMSV}(\xi)\rangle = \hat S_{AB}(\xi)|0,0\rangle$.

We choose the Gaussian unitary as a balanced beam-splitter operation, $\hat{U}_{\text{G}} = \hat{U}_{\text{BS}}(\pi/4)$. It gives
\begin{align}
\hat{U}_{\text{BS}} (\pi/4) |\text{TMSV} (\xi_j) \rangle &= |\text{SV} (\xi_j) \rangle \otimes |\text{SV}(-\xi_j) \rangle,
\end{align}
where $|\text{SV} (\xi_j) \rangle = \hat{S} (\xi_j) |0\rangle$ is the one-mode squeezed vacuum state, generated by the squeezing operator $\hat{S}(\xi_j)= \exp{\left[ 1/2\left( \xi_j \hat{a}^{\dagger 2} -\xi_j^* \hat{a}^2 \right)  \right] }$. Therefore, applying this unitary to the state $\rho_{\text{mix(TMSV)}}$ yields
\begin{align}
\hat{U}_{\text{BS}} (\pi/4) \rho_{\text{mix(TMSV)}} \hat{U}_{\text{BS}}^\dagger (\pi/4)  = p |\text{SV}(\xi_1) \rangle \langle \text{SV}(\xi_1) | \otimes  |\text{SV}(-\xi_1) \rangle \langle \text{SV}(-\xi_1) | + (1-p) |\text{SV}(\xi_2) \rangle \langle \text{SV}(\xi_2) | \otimes  |\text{SV}(-\xi_2) \rangle \langle \text{SV}(-\xi_2) |,
\end{align}
which is a separable state. This implies
\begin{align}
E_{\text{NG}} \left( \rho_{\text{mix(TMSV)}} \right) =1.
\end{align}

\subsubsection{c.\quad Photon-subtractions on a pure two-mode Gaussian state }

We show that the state generated by applying photon-subtractions on a pure two-mode Gaussian state is Gaussian-entanglable. The unnormalized state is expressed as
\begin{align}
| \tilde{\psi}_{\text{sub(k)}} \rangle &=  \hat{a}^k \hat{U}'_{\text{G}}  |0,0\rangle.
\end{align}
where $\hat{U}'_{\text{G}}$ is a two-mode Gaussian unitary. This state can be reexpressed by
\begin{align}
|\tilde{\psi}_{\text{sub(k)}} \rangle = \hat U'_{\text{G}} \left( \hat U_{\text{G}}^{'\dagger} \hat{a} \hat{U}'_{\text{G}} \right)^k |0,0\rangle .
\end{align}
Here, we can write $\hat{L}:=\hat{U}_{\text{G}}^{'\dagger} \hat{a} \hat{U}'_{\text{G}} = c_0 + c_A \hat{a} + c_B \hat{b} + d_A \hat{a}^{\dagger} + d_B \hat{b}^\dagger$. In this way, the resulting state can be further expressed as
\begin{align}
\hat{L}^k |0,0\rangle = \mathcal{P}^{(k)} \left( d_A \hat{a}^\dagger + d_B \hat{b}^\dagger \right) |0,0\rangle,
\end{align}
where $\mathcal{P}^{(k)}(x)$ denotes the polynomial at most $k$-th order of the variable $x$. 

We can prove that: For $k=0$, we have $\hat{L}^0 |0,0\rangle = \mathcal{P}^{(0)} \left( d_A \hat{a}^\dagger + d_B \hat{b}^\dagger \right) |0,0\rangle=|0,0\rangle$, thus obtain $\mathcal{P}^{(0)}=1$. Assume the statement $\hat{L}^k |0,0\rangle = \mathcal{P}^{(k)} \left( d_A \hat{a}^\dagger + d_B \hat{b}^\dagger \right)|0,0\rangle$ holds for $k\geq 0$. Thus for $k+1$, we can further write 
\begin{align}
\hat{L}^{k+1} |0,0\rangle &= \hat{L} \mathcal{P}^{(k)} \left( d_A \hat{a}^\dagger + d_B \hat{b}^\dagger \right)|0,0\rangle \nonumber \\
&= \left(  c_0 + c_A \hat{a} + c_B \hat{b} + d_A \hat{a}^{\dagger} + d_B \hat{b}^\dagger  \right)  \mathcal{P}^{(k)} \left( d_A \hat{a}^\dagger + d_B \hat{b}^\dagger \right)|0,0\rangle \nonumber \\
&=  \left( c_0 + d_A \hat{a}^{\dagger} + d_B \hat{b}^\dagger  \right) \mathcal{P}^{(k)} \left( d_A \hat{a}^\dagger + d_B \hat{b}^\dagger \right)|0,0\rangle 
+ \left( c_A \hat{a} + c_B \hat{b} \right)  \mathcal{P}^{(k)} \left( d_A \hat{a}^\dagger + d_B \hat{b}^\dagger \right)|0,0\rangle 
\nonumber \\
&=  \left( c_0 + d_A \hat{a}^{\dagger} + d_B \hat{b}^\dagger  \right) \mathcal{P}^{(k)} \left( d_A \hat{a}^\dagger + d_B \hat{b}^\dagger \right)|0,0\rangle 
+ \left[ \left( c_A \hat{a} + c_B \hat{b} \right),  \mathcal{P}^{(k)} \left( d_A \hat{a}^\dagger + d_B \hat{b}^\dagger \right) \right]|0,0\rangle 
\nonumber \\
&=  \left( c_0 + d_A \hat{a}^{\dagger} + d_B \hat{b}^\dagger  \right) \mathcal{P}^{(k)} \left( d_A \hat{a}^\dagger + d_B \hat{b}^\dagger \right)|0,0\rangle 
+  \left( c_A d_A + c_B d_B \right) \left(\mathcal{P}^{(k)} \right)' \left( d_A \hat{a}^\dagger + d_B \hat{b}^\dagger \right) |0,0\rangle 
\nonumber \\
&=  \mathcal{P}^{(k+1)} \left( d_A \hat{a}^\dagger + d_B \hat{b}^\dagger \right)|0,0\rangle.
\end{align}
In this derivation, we use the conditions $\left( c_A \hat{a} +c_B \hat{b} \right)|0,0\rangle = 0$ and $\left[ c_A \hat{a} +c_B \hat{b}, d_A \hat{a}^\dagger + d_B \hat{b}^\dagger \right] = c_A d_A + c_B d_B$.

A passive Gaussian unitary $\hat{V}$ can rotate the state, so that,
\begin{align}
\hat{V} \left( d_A \hat{a}^\dagger + d_B \hat{b}^\dagger \right) \hat{V}^\dagger = d \hat{a}^\dagger,
\end{align}
where $d=\sqrt{ |d_A|^2 + |d_B|^2 }$. Consequently, we have
\begin{align}
\hat{V} \hat{L}^k |0,0\rangle = |\psi_k \rangle_A \otimes |0\rangle_B ,
\end{align}
which is a product state. $|\psi_k \rangle_A = \sum_{n=0}^k c_n |n\rangle_A$ is a single-mode states.

Finally, if we choose the Gaussian unitary as $\hat{U}_{\text{G}} = \hat{V} \hat{U}_{\text{G}}^{'\dagger}$, we will have the product state,
\begin{align}
\hat{U}_{\text{G}} | \tilde{\psi}_{\text{sub(k)}} \rangle &= |\psi_k \rangle_A \otimes |0\rangle_B.
\end{align}
This implies that 
\begin{align}
E_{\text{NG}}(|\psi_{\text{sub(k)}}\rangle)=1 .
\end{align}

\section{II.\quad Entanglement distillation of non-Gaussian entangled state in Gaussian settings}

It is proved that Gaussian entangled states cannot be distilled by the protocols consisting of arbitrary local Gaussian unitaries, local homodyne detections, and classical communication~\cite{EisertPRL2002}. Here we show that the non-Gaussian entanglement, certified by the witness $E_{\rm NG}(\rho)>1$ can serve as an effective resource for entanglement distillation in such protocol. As an example, we use the two-mode squeezed Kerr states as the input in a two-copy distillation protocol.

We consider two identical states $\rho$ are prepared on four modes A, B, C, D as $\rho_{ABCD} = \rho_{AC} \otimes \rho_{BD}$. Here Alice holds modes A and B, while Bob holds modes C and D. 
Then they locally apply identical two-mode Gaussian unitaries $\hat U_{AB}=\hat U_{CD} \equiv \hat U$ on the bipartition $AB|CD$, which yields $\rho'_{ABCD} = \left( \hat{U}_{AB} \otimes \hat{U}_{CD} \right) \rho_{ABCD} \left( \hat{U}^\dagger_{AB} \otimes \hat{U}^\dagger_{CD} \right)$. We consider the Gaussian unitaries in the entanglement-generating form~\cite{MattiaArXiv2026}, $\hat{U} (\theta_1,\theta_2,\xi)=\hat{U}_{\rm BS} (\theta_1) \hat{S} (\xi) \hat{U}_{\rm BS}^{\dagger} (\theta_2) $, where $\hat{U}_{\rm BS}(\theta_j) = e^{\theta_j \left( \hat{a}^\dagger \hat{b} -\hat{a} \hat{b}^\dagger \right)}$ is the beam-splitter operation and $\hat{S}(\xi) = e^{\xi \hat{a}^\dagger \hat{b}^\dagger - \xi^* \hat{a}\hat{b}}$ is the two-mode squeezing operation.
Finally, Alice and Bob perform homodyne measurements on modes B and D, respectively. The measured observable is $\hat{M}(\phi) = \cos{(\phi_M)} \hat{X} + \sin{(\phi_M)} \hat{P} $. If we choose the equal measurement outcomes $m_B=m_D=m$ from both parties, and denote $|m \rangle$ the corresponding measurement basis, we obtain the conditional state on modes $A,C$ as $\rho_{\rm d}$.
We optimize over the Gaussian unitary parameters and the homodyne detections, the entanglement distillation is successfully achieved if the conditional output state has larger entanglement than the input state, $\max_{\theta_1, \theta_2,\xi, \phi_{\rm M}, m } E(\rho_{\rm d}) > E(\rho) $.
In the main text, this protocol is applied to lossy two-mode squeezed Kerr states, showing that significant distillation gain is reached when the non-Gaussian entanglement witness $E_{\rm NG}>1$ is clearly observed.

\section{III.\quad NOON-type non-Gaussian entanglement witness}

\subsection{A.\quad The threshold $f_{\text{G}}$ is determined by the maximum Schmidt coefficient}

Our NOON-type non-Gaussian entanglement witness is
\begin{align}
f = \text{Tr}[\mathcal{F} \rho],
\end{align}
with the operator
\begin{align}\label{SMeq:N00Nwitness}
\mathcal{F} &= \frac{1}{2} \left( |0,N \rangle \langle 0, N| + |N,0 \rangle \langle N,0| + |0,N \rangle \langle N, 0| +|N,0 \rangle \langle 0,N| \right).
\end{align}
We define $f_{\text{G}}$ as the maximum value calculated by optimizing any two-mode Gaussian unitaries on pure product states,
\begin{align}
f_{\text{G}} &:= \max_{\hat{U}_{\text{G}} \in \mathcal{G}_2, |\psi_A\rangle, |\psi_B \rangle } \langle \psi_A | \otimes \langle \psi_B | \hat{U}_{\text{G}}^\dagger \mathcal{F} \hat{U}_{\text{G}} |\psi_A \rangle \otimes |\psi_B \rangle \nonumber \\
&= \max_{\hat{U}_{\text{G}} \in \mathcal{G}_2} \max_{ |\psi_A\rangle, |\psi_B \rangle }  \left\vert \langle \psi_{\text{NOON(G)}} |\psi_A \rangle \otimes |\psi_B \rangle \right\vert^2
\end{align}
where $|\psi_{\text{NOON(G)}} \rangle = \hat{U}_{\text{G}}^\dagger |\psi_{\text{NOON}} \rangle$. By Schmidt decompositions, we have $|\psi_{\text{NOON(G)}} \rangle = \sum_i \lambda_i |n_i \rangle |m_i \rangle$,  with $\lambda_i > 0$ and $\sum_i \lambda_i^2 =1$. We denote $\lambda_{\max}$ as the maximum Schmidt coefficient. 
As proved in Ref~\cite{BourennanePRL2004}, the maximum overlap between a pure state $\ket{\psi}$ and any pure product state is equivalent to the maximal Schmidt coefficient of $\ket{\psi}$. We first calculate
\begin{align}
\left\vert \langle \psi_{\text{NOON(G)}} |\psi_A \rangle \otimes |\psi_B \rangle \right\vert &= \left\vert \sum_i \lambda_i \langle n_i |\psi_A \rangle \langle m_i | \psi_B \rangle  \right \vert \nonumber \\
&\leq  \sum_i \left\vert \lambda_i \langle n_i |\psi_A \rangle \langle m_i | \psi_B \rangle  \right \vert \nonumber \\
&=  \sum_i  \lambda_i \left\vert \langle n_i |\psi_A \rangle \right \vert  \left\vert \langle m_i | \psi_B \rangle  \right \vert \nonumber \\
& \leq \lambda_{\max} \sum_i \left \vert \langle n_i |\psi_A \rangle \right \vert \left \vert \langle m_i | \psi_B \rangle  \right \vert.
\end{align}
Considering $\sum_i \left \vert \langle n_i |\psi_A \rangle \right \vert^2 \leq 1,  \sum_i \left \vert \langle m_i |\psi_B \rangle \right \vert^2 \leq 1$ and the Cauchy-Schwarz inequality
\begin{align}
\sum_i \left \vert \langle n_i |\psi_A \rangle \right \vert \cdot \left \vert \langle m_i | \psi_B \rangle  \right \vert \leq \sqrt{ \sum_i \left \vert \langle n_i |\psi_A \rangle \right \vert^2 } \sqrt{ \sum_i \left \vert \langle m_i |\psi_B \rangle \right \vert^2 } \leq 1.
\end{align}
Therefore we obtain
\begin{align}
\left\vert \langle \psi_{\text{NOON(G)}} |\psi_A \rangle \otimes |\psi_B \rangle \right\vert \leq \lambda_{\max}.
\end{align}
The inequality is saturated when $|\psi_A\rangle = |n_{i^*} \rangle,  |\psi_B\rangle = |m_{i^*} \rangle$, with $\lambda_{i^*} = \lambda_{\max}$. In this way, we obtain
\begin{align}
f_{\text{G}} = \max_{\hat{U}_{\text{G}} \in \mathcal{G}_2} \lambda_{\max}^2 \left(  |\psi_{\text{NOON(G)}} \rangle \right).
\end{align}
The genuine non-Gaussian entanglement of $\rho$ is revealed if this GE threshold is surpassed, $f > f_{\text{G}}$.

The optimization in the NOON-type witness can be simplified in the similar way as for the witness $E_{\rm NG}$ in SM Sec. I. B, while using the fact that Schmidt coefficients are invariant under local unitaries. Therefore, based on the effective Gaussian unitary in Eq.~\eqref{SMeq:UGeff},
\begin{align}
\hat{U}_{\text{G}}^{\rm eff} 
&= \hat{U}_{\text{BS}} (\theta^\text{II}) \left[ \hat{R}_A(\theta^{\text{II}}_0) \hat{S}_A(r) \hat{R}_A(\theta^{\text{I}}_1) \otimes  \hat{R}_B (\theta^{\text{I}}_2) \right] \hat{U}_{\text{BS}} (\theta^{\text{I}}) \hat{R}_A (\theta^{\text{I}}_0), 
\end{align}
the threshold for the NOON-type witness can be evaluated as 
\begin{align}
f_{\rm G} = \max_{ \hat{U}_{\rm G}^{\rm eff} \in \mathcal{G}_2^{\rm eff} } \lambda_{\max}^2 \left(\hat{U}_{\rm G}^{\rm eff} |\psi_{\rm NOON} \rangle  \right).
\end{align}

\subsection{B.\quad Bounds on $f_{\text{G}}$ }
For any Gaussian unitary and any product state, the NOON-type witness satisfies
\begin{align}
\langle \psi_A  \otimes \psi_B |\hat{U}_{\text{G}}^\dagger \mathcal{F} \hat{U}_{\text{G}} | \psi_A  \otimes \psi_B \rangle = \left \vert \langle \psi_{\text{NOON(G)}}| \psi_A \otimes \psi_B \rangle \right \vert^2 \leq 1,
\end{align}
which gives $f_{\text{G}} \leq 1$.

The lower bound of $f_{\text{G}}$ can be obtained by choosing some specific Gaussian unitaries. First, by choosing the identity operator $\hat{U}_{\text{G}} = \hat{I}$, we obtain 
\begin{align}
\hat{U}_{\text{G}} |\psi_{\text{NOON}} \rangle = \frac{1}{\sqrt{2}} \left( |N,0 \rangle + |0,N\rangle \right),
\end{align}
whose maximum Schmidt coefficient is $\lambda_{\max} = \frac{1}{\sqrt{2}}$. This yields
\begin{align}
f_{\text{G}} \geq \lambda_{\max}^2 \left( \hat{I} |\psi_{\text{NOON}} \rangle \right) = \frac{1}{2}.
\end{align}
A second lower bound is obtained from a balanced beam splitter together with a local phase rotation.  
We choose $\hat{U}_{\text{G}} = \hat{U}_{\text{BS}} \hat{R}_A(\phi)$, where $\hat{R}_A(\phi) = e^{-i\phi \hat{a}^\dagger \hat{a}}$ is a local rotation operator and $\hat{U}_{\text{BS}}(\pi/4) =\exp{\left[ \pi/4 \left(\hat{a}^\dagger \hat{b} - \hat{a} \hat{b}^\dagger \right) \right]}$ is a balanced beam-splitter. This gives
\begin{align}
\hat{U}_{\text{G}} |\psi_{\text{NOON}} \rangle &= 2^{-(N+1)/2} \sum_{j=0}^N \sqrt{\binom{N}{j} } \left[ e^{-iN\phi} (-1)^{N-j} +1 \right] |j, N-j \rangle.
\end{align}
One can always find an optimized phase shift $\phi$, so that $e^{-iN\phi} (-1)^{N-j'}=1$, where $j'=\lfloor N/2\rfloor$ corresponds to the maximal binomial coefficient. In this way, the maximum Schmidt coefficient is 
\begin{align}
\lambda_{\max} = 2^{-(N-1)/2}\sqrt{\binom{N}{\lfloor N/2\rfloor}},
\end{align}
yielding
\begin{align}
f_{\text{G}} \geq \lambda_{\max}^2 \left( \hat{U}_{\text{BS}} \hat{R}_A(\phi) |\psi_{\text{NOON}} \rangle \right) = 2^{1-N} \binom{N}{\lfloor N/2\rfloor} .
\end{align}

Finally, we obtain the inequalities
\begin{align}
\max \left\{ \frac{1}{2}, 2^{1-N} \binom{N }{\lfloor N/2 \rfloor } \right\}   \leq f_{\text{G}} \leq 1.
\end{align}

\subsection{C.\quad  Certifying genuine non-Gaussian entanglement with the GE threshold $f_{\text{G}}$}

We show that the GE threshold $f_{\text{G}}$, although defined by optimizing over pure GE states, is also the maximum value attainable for any states $\rho\in \mathcal{GE}_{\text{mix}}$. Hence, $f>f_{\text{G}}$ certifies genuine non-Gaussian entanglement. 

For pure GE states, the threshold is expressed
\begin{align}
f_{\text{G}}^{\rm pure} =f_{\text{G}} = \max_{\hat{U}_{\text{G}} \in \mathcal{G}_2, |\psi_A\rangle, |\psi_B \rangle } \langle \psi_A | \otimes \langle \psi_B | \hat{U}_{\text{G}}^\dagger \mathcal{F} \hat{U}_{\text{G}} |\psi_A \rangle \otimes |\psi_B \rangle.
\end{align}
Analogously, the bound based on separable states $\rho_{\text{sep}} = \sum_i c_i |\psi_A^{(i)} \rangle \langle \psi_A^{(i)}| \otimes |\psi_B^{(i)}\rangle \langle \psi_B^{(i)}|$, where $c_i\geq0, \sum_i c_i =1$, can be expressed as
\begin{align}
f_{\text{G}}^{\rm sep} &= \max_{\hat{U}_{\text{G}} \in \mathcal{G}_2, \rho_{\text{sep}} } \text{Tr} \left( \mathcal{F} \hat{U}_{\text{G}} \rho_{\text{sep}} \hat{U}_{\text{G}}^\dagger \right) \nonumber \\
&= \max_{\hat{U}_{\text{G}} \in \mathcal{G}_2, c_i,|\psi_A \rangle^{(i)}, |\psi_B\rangle^{(i)} } \sum_i c_i \langle \psi_A^{(i)} \otimes \psi_B^{(i)} | \hat{U}_{\text{G}}^\dagger \mathcal{F} \hat{U}_{\text{G}} | \psi_A^{(i)} \otimes \psi_B^{(i)} \rangle. 
\end{align}
For each term $\langle \psi_A^{(i)} \otimes \psi_B^{(i)} | \hat{U}_{\text{G}}^\dagger \mathcal{F} \hat{U}_{\text{G}} | \psi_A^{(i)} \otimes \psi_B^{(i)} \rangle \leq f_{\text{G}}^{\rm pure} = f_{\rm G}$, we obtain $f_{\text{G}}^{\rm sep} \leq f_{\text{G}}$. Since pure product states are a specific case of separable states, this gives $f_{\text{G}}^{\rm sep} = f_{\text{G}}$.

Then, we consider the states from the set $\mathcal{GE}_{\text{mix}}$, the corresponding bound is 
\begin{align}
f_{\text{G}}^{\rm mix} &= 
 \max_{\hat{U}^{(j)}_{\text{G}} \in \mathcal{G}_2, \rho^{(j)}_{\text{sep}}  } \sum_j p_j \text{Tr} \left( \mathcal{F} \hat{U}^{(j)}_{\text{G}} \rho^{(j)}_{\text{sep}} \hat{U}_{\text{G}}^{(j)\dagger} \right). 
\end{align}
Since it is known that each term satisfies $\text{Tr} \left( \mathcal{F} \hat{U}^{(j)}_{\text{G}} \rho^{(j)}_{\text{sep}} \hat{U}_{\text{G}}^{(j)\dagger} \right) \leq f_{\text{G}}$, we can also obtain $f_{\text{G}}^{\rm mix} \leq f_{\text{G}}$. Pure GE states are involved in the set $\mathcal{GE}_{\rm mix}$, thus we obtain $f_{\rm G}^{\rm mix}=f_{\rm G}$.

Therefore, if any state $\rho$ satisfies
\begin{align}
f=\text{Tr}(\mathcal F\rho)>f_{\rm G},
\end{align}
then $\rho\notin\mathcal{GE}_{\rm mix}$, certifying the genuine non-Gaussian entanglement.

\end{widetext}

\end{document}